\documentclass[aps, a4paper,superscriptaddress,12pt,preprintnumbers,floatfix,nofootinbib]{revtex4}
\usepackage{url}
\usepackage{xcolor}
\usepackage{hyperref}
\usepackage[normalem]{ulem}
\hypersetup{colorlinks=true,linkcolor=redLinks,citecolor=greenLinks,
urlcolor=redLinks,
pdfborder={0 0 1}}
\hypersetup{ 
    bookmarks=true,         
    unicode=false,          
    pdftoolbar=true,        
    pdfmenubar=true,        
    pdffitwindow=false,     
    pdfstartview={FitH},    
    pdftitle={My title},    
    pdfauthor={Author},     
    pdfsubject={Subject},   
    pdfcreator={Creator},   
    pdfproducer={Producer}, 
    pdfkeywords={keyword1, key2, key3}, 
    pdfnewwindow=true,      
    colorlinks=false,       
    linkcolor=red,          
    citecolor=green,        
    filecolor=magenta,      
    urlcolor=cyan           
}
\colorlet{shadecolor}{gray!15}
\definecolor{greenLinks}{rgb}{0, 0.6, 0} 
\definecolor{blueLinks}{rgb}{0, 0, 0.6}
\definecolor{redLinks}{rgb}{0.6, 0, 0}
\definecolor{tempText}{rgb}{0.55, 0.10,0.67}
\definecolor{eprintLinks}{rgb}{0.4, 0.4, 0.4}
\definecolor{journalLinks}{rgb}{0.6, 0, 0}
\usepackage{subfigure}
\usepackage{rotating}
\usepackage{slashed}     

\usepackage[utf8]{inputenc}
\usepackage[english]{babel}
\usepackage{amsmath}
\usepackage{graphicx}
\usepackage{color}
\usepackage{mathrsfs}   
\usepackage{amssymb}
\usepackage{hyperref}
\usepackage{dcolumn}
\usepackage{bm}
\usepackage{graphicx}

\newcommand {\ignore}[1]{}
\def\vev#1{\left\langle #1\right\rangle}
\newcommand{\sm}{Standard Model }
\def\SM{$\mathrm{SU(3)_c \otimes SU(2)_L \otimes U(1)_Y}$ }
 \def\znbb {neutrinoless double beta decay }
\def\TrTrOne{ $\mathrm{SU(3)_c \otimes SU(3)_L \otimes U(1)_X}$ }
 \def\one{\ensuremath{\mathbf{1}}}
 \def\three{\ensuremath{\mathbf{3}}}
 \def\threeS{\ensuremath{\mathbf{3^*}}}

\def\beqn#1{\begin{equation}\label{#1}}
\def\eeqn{\end{equation}}

\def\beqa#1{\begin{eqnarray}\label{#1}}
\def\eeqa{\end{eqnarray}}

\newcommand{\AddrAHEP}{
  {\it AHEP Group, Instituto de F\'{\i}sica Corpuscular --
    C.S.I.C./Universitat de Val{\`e}ncia \\
    Edificio de Institutos de Paterna,
 C/Catedratico Jos\'e Beltr\'an, 2 \\E-46980 Paterna (Val\`{e}ncia) - SPAIN}}

\begin{document}
\title{Realistic \TrTrOne model with a type II  Dirac neutrino seesaw mechanism }
\author{Mario Reig}
\email{mareiglo@alumni.uv.es}
\affiliation{\AddrAHEP}
\author{Jos\'e W.F. Valle}
\email{valle@ific.uv.es}
\affiliation{\AddrAHEP}
\author{C.A. Vaquera-Araujo}
\email{vaquera@ific.uv.es}
\affiliation{\AddrAHEP}
  \vspace{1cm}

\begin{abstract}
   \vspace{1cm}

  Here we propose a realistic \TrTrOne electroweak gauge model with
   enlarged Higgs sector.  The scheme allows for the natural
   implementation of a type II seesaw mechanism for Dirac neutrinos,
   while charged lepton and quark masses are reproduced in a natural way thanks to the presence of new scalars. The new \TrTrOne energy
   scale characterizing neutrino mass generation could be accessible
   to the current LHC experiments.

\end{abstract}

\maketitle

\section{Introduction}

Despite the fierce experimental effort over the
last decades the long-standing 
challenge concerning the question of whether neutrinos are their own
anti-particles
remains~\cite{Barabash:2011fn,Avignone:2007fu,Blot:2016cei}.
Although neutrinos could be Dirac or Majorana fermions, the leading
theoretical expectation is that they are Majorana, the general belief
being that the smallness of neutrino masses relative to the other \sm
fermion masses is due to their charge neutrality.
This fits naturally to the idea that neutrinos acquire Majorana masses
from Weinberg's dimension five operator. Realizations include various
varieties of
type I~\cite{gell-mann:1980vs,yanagida:1979,mohapatra:1980ia,Schechter:1980gr,Schechter:1981cv}
or type II~\cite{Schechter:1980gr,Schechter:1981cv,Lazarides:1980nt,Grimus:2009mm}
seesaw mechanisms, irrespective of whether the seesaw is realized at
high or at low mass scale~\cite{Boucenna:2014zba}.
Until the observation of \znbb~\cite{klapdor-kleingrothaus:2004wj}
becomes a reality~\cite{Barabash:2011fn} we must keep an eye open to
the possibility that neutrinos can be Dirac particles as well.

There are two issues that Dirac neutrino models must face, namely
predicting the Dirac nature of neutrinos, and understanding their
small mass.
Dirac neutrinos require extra symmetries beyond \SM gauge symmetry,
otherwise massive neutrinos are generally expected to be Majorana
fermions. Within the standard \SM electroweak gauge structure this can
be ensured by imposing a conserved lepton numberlike symmetry,
continuous~\cite{peltoniemi:1993ss} or discrete~\cite{Chulia:2016ngi}.
Likewise, one may consider schemes based on flavor symmetries, as
suggested in~\cite{Aranda:2013gga} or appealing to the existence of
extra dimensions~\cite{arkani-hamed:1998vp,Chen:2015jta}.
Alternatively one may extend the gauge group, so as to include the
lepton number symmetry~\cite{Ma:2015mjd}, for example, by using the
extended \TrTrOne gauge structure~\cite{Singer:1980sw} although most
formulations lead to Majorana
neutrinos~\cite{Tully:2000kk,Kitabayashi:2000nq,Kitabayashi:2001dg,Montero:2001ts,Chang:2006aa,Dong:2006mt,Mizukoshi:2010ky,Boucenna:2014ela,Boucenna:2014dia,Okada:2015bxa,Boucenna:2015zwa,Vien:2016tmh}.

Because of the special features of the \TrTrOne based models with
respect to other electroweak extensions based, for example, on
left-right symmetry, in this paper we focus on the possibility of
having naturally light Dirac neutrinos with seesaw-induced masses,
within the  \TrTrOne gauge structure.
However, in contrast to Ref.~\cite{Valle:2016kyz} in order to predict
realistic quark masses in a natural way, two extra antitriplet scalar
multiplets are included.
As before, we have a lepton numberlike symmetry $\mathcal{L}$,
preserved in the leptonic and quark sector.
This symmetry is softly broken in the scalar sector by the term
$f\phi_0 \phi_1 \phi_2$ so that in the limit $f \to 0$ the Lagrangian
symmetry gets enhanced.
We show that the smallness of $f$ is related to the smallness of
neutrino mass.
This way we recover the new variant of type II seesaw mechanism for
Dirac neutrinos recently considered in~\cite{Valle:2016kyz}.
However, in contrast to the previous Ref.~\cite{Valle:2016kyz}, here
the naturally small induced vacuum expectation values (vevs)
responsible for neutrino mass generation are decoupled from the quark
sector.
This eliminates the need of fine-tuning the Yukawa couplings so that
all fermion masses are naturally reproduced in a realistic way, since exotic quarks and standard model quarks are now decoupled.
The scales associated to neutrino mass generation and the new \TrTrOne
gauge interactions could be accessible to the current LHC experiments.

\section{Model}
\label{model}

The model is a modified version of the one presented in
Ref.~\cite{Valle:2016kyz}. In the new setup, two extra scalar triplets
are included and an auxiliary $\mathbb{Z}_4$ symmetry is implemented
in order to decouple heavy quarks. The matter content and the
transformation properties of the fields are contained in Table
\ref{tab:content2}

\begin{table}[!h]
\begin{center}
\begin{tabular}{|c||c|c|c|c||c|c|c|c|c|c||c|c|c|c|c|}
\hline
 & $\psi_L^\ell$ & $\ell_R$ & $S_R^{\ell}$ & $\tilde{S}_R^{\ell}$ & $Q_L^{1,2}$  & $Q_L^3$ & $\hat{u}_{R}$ & $U_{R}$ & $\hat{d}_{R}$& $D_{R}^{1,2}$  & $\phi_0$ & $\phi_1$ & $\phi_2$ & $\phi_3$&$\phi_4$\\
\hline
\hline
$\mathrm{SU(3)_c}$ & \one &\one &\one &\one &\three &\three &\three&\three &\three &\three &\one&\one &\one &\one&\one\\
\hline
$\mathrm{SU(3)_L}$ & \threeS & \one & \one &\one &\three & \threeS & \one & \one & \one & \one & \threeS & \threeS& \threeS  &\threeS &\threeS  \\
\hline
$\mathrm{U(1)_X}$ & $-\frac{1}{3}$ & $-1$ &  $0$ &  $0$ & $0$ & $+\frac{1}{3}$ & $+\frac{2}{3}$ & $+\frac{2}{3}$ & $-\frac{1}{3}$ & $-\frac{1}{3}$ & $+\frac{2}{3}$& $-\frac{1}{3}$& $-\frac{1}{3}$& $-\frac{1}{3}$&$-\frac{1}{3}$\\ 
\hline
$\mathcal{L}$ &$-\frac{1}{3}$&$-1$ &$+1$ &$+1$ & $ -\frac{2}{3}$& $+\frac{2}{3}$ & $0$& $1$&$0$ &$-1$ & $+\frac{2}{3}$ &$-\frac{4}{3}$ & $-\frac{4}{3}$& $-\frac{1}{3}$& $+\frac{2}{3}$\\
\hline
$ \mathbb{Z}_3 $ & $\omega$ & $\omega$ &  $\omega$&  $\omega$ & $\omega ^{2}$ & $\omega ^{2}$ & $\omega ^{2}$& $\omega ^{2}$ & $\omega ^{2}$& $\omega ^{2}$ & $1$ & $1$ & $1$ & $1$ & $1$ \\ 
\hline
$ \mathbb{Z}_4 $ & $1$ & $1$ &  $-i$&  $i$ & $1$ & $1$ & $1$& $-1$ & $1$& $-1$ &$1$ & $i$&$-i$& $-1$ & $1$\\ 
\hline
\end{tabular}\caption{Matter content of the model, where $\ell=(e,\mu,\tau)$,
  $\hat{u}\equiv{(u,c,t)}$ and $\hat{d}\equiv{(d,s,b)}$.} 
\label{tab:content2}
\end{center}
\end{table}
In terms of the gauge group generators, the electric charge is expressed as
\begin{equation}
Q=T_3 +\frac{1}{\sqrt{3}}T_8 +X\,,
\end{equation}
while lepton number is defined as
\begin{equation}
L=\frac{4}{\sqrt{3}}T_8 +\mathcal{L}\,.
\end{equation}

The symmetry breaking pattern is assumed to be of the form
\begin{gather}\label{scalar3plets2}
\vev{\phi_0}=\frac{1}{\sqrt{2}}\left(
\begin{array}{c}
k_0\\
0\\
0
\end{array}
\right), \quad
\vev{\phi_1}=\frac{1}{\sqrt{2}}\left(
\begin{array}{c}
0\\
\delta_1\\
n_1
\end{array}
\right),\quad
\vev{\phi_2}=\frac{1}{\sqrt{2}}\left(
\begin{array}{c}
0\\
\delta_2\\
n_2
\end{array}
\right),\\
\vev{\phi_3}=\frac{1}{\sqrt{2}}\left(
\begin{array}{c}
0\\
0\\
n_3
\end{array}
\right),\quad
\vev{\phi_4}=\frac{1}{\sqrt{2}}\left(
\begin{array}{c}
0\\
k_4\\
0
\end{array}
\right),
\end{gather}
with $n_{1,2,3}\gg k_{0,4}\,,\delta_{1,2}$ and $\delta_1\sim\delta_2$.

The Yukawa interactions invariant under all the defining symmetries of
the model are
\begin{eqnarray}\label{lagY2}
 -\mathcal{L}_{\rm f} &=& \,
 y^{\ell} \bar{\psi}_L \,\ell_R \,\phi_0 + y_1 \,\bar{\psi}_L \, S_{R} \,\phi_1  + \tilde{y}_{2} \,\bar{\psi}_L \, \tilde{S}_{R} \,\phi_2 \nonumber\\
 &+&  
y^{u}\, \bar{Q}_L^{1,2} \,\hat{u}_R \,\phi_0^*  +  
\hat{y}^{u}\, \bar{Q}_L^{3} \,\hat{u}_R \,\phi_4 
+  y^{U}\, \bar{Q}_L^{3} \,U_R \,\phi_3 \nonumber \\
&+& 
y^{d} \,\bar{Q}_L^{3} \,\hat{d}_R \,\phi_0 + 
\hat{y}^{d} \, \bar{Q}_L^{1,2} \,\hat{d}_R\,\phi_4^* 
  +  y^{D} \, \bar{Q}_L^{1,2} \,D_R^{1,2} \,\phi_3^* 
 + \mathrm{h.c.}
\end{eqnarray}
In this setup, $\phi_1$ and $\phi_2$ are completely decoupled from the
quark sector and are responsible for the neutrino mass generation,
whereas $\phi_3$ and $\phi_4$ contribute exclusively to the quark
masses.  After spontaneous symmetry breaking, the above interactions
lead to the following mass matrices for the fermion fields:
\begin{itemize}
\item charged leptons: $m_\ell= \frac{k_0}{\sqrt{2}} y^{\ell}\,, $
\item neutrinos:
\begin{equation}
\begin{split}
m_{\nu}&=\frac{1}{\sqrt{2}}\left(\begin{array}{cc}
 \delta_1 y_{1} & \delta_2 \tilde{y}_{2}\\
 n_1 y_{1} & n_2  \tilde{y}_{2}\\
\end{array}\right)=\frac{1}{\sqrt{2}}\left(
\begin{array}{cc}
 \delta_1 \mathbb{I} & \delta_2 \mathbb{I} \\
 n_1 \mathbb{I} & n_2 \mathbb{I} \\
\end{array}
\right)\left(
\begin{array}{cc}
 y_1 & 0 \\
 0 & \tilde{y}_2 \\
\end{array}
\right)\,,
\end{split}
\end{equation}
\item up-type quarks, basis $(u,c,t,U)$:
\begin{equation}
\begin{split}
m_{u}&=\frac{1}{\sqrt{2}}\left(
\begin{array}{cccc}
 k_0 y^{u}_1 & k_0 y^{c}_1 & k_0 y^{t}_1 & 0 \\
 k_0 y^{u}_2 & k_0 y^{c}_2 & k_0 y^{t}_2 & 0 \\
 k_4 \hat{y}^{u} & k_4 \hat{y}^{c} & k_4 \hat{y}^{t} & 0 \\
0 & 0 & 0 & n_3 y^{U} \\
\end{array}
\right)\,,
\end{split}
\end{equation}
\item down-type quarks, basis $(d,s,b,D^{1},D^{2})$:
\begin{equation}
\begin{split}
m_{d}&=\frac{1}{\sqrt{2}}\left(
\begin{array}{ccccc}
 k_4\hat{y}_1^{d} & k_4 \hat{y}_1^{s} &  k_4 \hat{y}_1^{b} & 0 & 0 \\
 k_4\hat{y}_2^{d} & k_4\hat{y}_2^{s} &  k_4 \hat{y}_2^{b} & 0 & 0 \\
 k_0 y^{d}  & k_0 y^{s}  & k_0 y^{b} & 0 & 0 \\
0 & 0 & 0 &  n_3 y_1^{D^{1}} & n_3 y_1^{D^{2}} \\
0 & 0 & 0 &  n_3 y_2^{D^{1}} &  n_3 y_2^{D^{2}} \\
\end{array}
\right)\,.
\end{split}
\end{equation}
\end{itemize}
Realistic quark masses can be readily obtained as the standard model and exotic
sectors are independent by virtue of the $\mathbb{Z}_4$ symmetry. This
also implies that the Cabibbo-Kobayashi-Maskawa matrix describing quark mixing is unitary.

Moreover, from the diagonalization of $m_{\nu}m_{\nu}^{\dagger}$, the
mass of the light neutrino in the one-family approximation is given by
\begin{equation}
m_{light}\approx\frac{1}{\sqrt{2}}\frac{|y_1 \tilde{y}_2 (\delta_2 n_1 - \delta_1 n_2)|}{\sqrt{n_1^{2}y_1^{2}+n_2^{2}\tilde{y}_2^{2}}}.
\end{equation}
Notice that the Dirac nature of neutrinos is ensured by the discrete
$ \mathbb{Z}_3 $ group.  In what follows, we specialize to the case
$\delta_1=\delta_2\equiv \delta$ and show that the smallness of the
neutrino mass can be understood as emerging from a type II seesaw
mechanism for Dirac neutrinos. This links the emergence of the small
induced scale $\delta$ to the breaking of the global symmetry
$\mathcal{L}$ \cite{Valle:2016kyz}.

The scalar potential of the model is
\begin{equation}
\begin{split}
V=&\sum_{i=0}^{4}\big ( \mu_{i}^{2}|\phi _{i}|^{2}+\lambda_{i}|\phi _{i}|^{4}\big )+\sum _{i< j}\big [ \lambda_{ij}|\phi _{i}|^{2}|\phi _{j}|^{2} +\tilde{\lambda}_{ij}(\phi^{\dagger}_{i}\phi_{j})(\phi^{\dagger}_{j}\phi_{i})\big ] \\ &
+ \big [ \lambda(\phi^{\dagger}_{1}\phi_{2})(\phi^{\dagger}_{1}\phi_{2})+f\phi_{0}\phi_{1}\phi_{2}+\mathrm{h.c.}\big ],
\end{split}
\end{equation}
where the term $f\phi_{0}\phi_{1}\phi_{2}$ breaks explicitly the
$\mathcal{L}$ symmetry. Assuming real vevs and parameters in the
potential, the tadpole equations can be solved for the parameters
$\mu^2_i$, $\tilde{\lambda}_{13}$, $\tilde{\lambda}_{14}$ and $f$ as
follows
\begin{eqnarray}
\mu_0^2&=&-\frac{1}{2}\left[2 \lambda _0 k_0^2+\lambda _{01} \left(\delta ^2+n_1^2\right)+\lambda _{02} \left(\delta ^2+n_2^2\right)+n_3^2 \lambda
   _{03}+k_4^2 \lambda _{04}\right]\nonumber\\
   &&-\frac{\delta ^2 \left(n_1-n_2\right)}{2 k_0^2}\left[\left(n_1-n_2\right) \left(\tilde{\lambda} _{12}+2 \lambda \right)+\frac{ n_2 \left(n_3^2 \tilde{\lambda} _{23}-k_4^2 \tilde{\lambda} _{24}\right)}{ \left(\delta ^2+n_1
   n_2\right)}\right]\,,\\
\mu_1^2&=&-\frac{1}{2}\left[k_0^2 \lambda _{01}+n_3^2 \lambda _{13}+k_4^2 \lambda _{14}+2  \left(\delta ^2+n_1^2\right)\lambda _1+\left(\delta ^2+n_2^2\right) \left(\lambda _{12}+\tilde{\lambda} _{12}+2 \lambda \right)\right]\nonumber\\
&&+\frac{n_2 \left(\delta ^2 k_4^2 \tilde{\lambda} _{24}+n_1 n_2 n_3^2 \tilde{\lambda} _{23}\right)}{2 n_1 \left(\delta ^2+n_1 n_2\right)}\,,\\
\mu_2^2&=&-\frac{1}{2}\left[k_0^2 \lambda _{02}+k_4^2 \lambda _{24}+2 \left(\delta ^2+n_2^2\right)\lambda _2 +\lambda _{23} n_3^2+\left(2 \lambda +\lambda _{12}+\tilde{\lambda} _{12}\right) \left(\delta ^2+n_1^2\right)\right]\nonumber\\
&&-\frac{\delta ^2 k_4^2 \tilde{\lambda} _{24}+n_1 n_2 n_3^2 \tilde{\lambda} _{23}}{2 \left(\delta ^2+n_1 n_2\right)}\,,\\
\mu_3^2&=&-\frac{1}{2}\big[k_0^2 \lambda _{03}+k_4^2 \lambda _{34}+\lambda _{13} \left(\delta ^2+n_1^2\right)+\lambda _{23} \left(\delta
   ^2+n_2^2\right)+2 \lambda _3 n_3^2\nonumber\\&&\qquad+n_2 \left(n_2-n_1\right) \tilde{\lambda} _{23}\big]\,,\\
\mu_4^2&=&-\frac{1}{2}\big[k_0^2 \lambda _{0,4}+2 k_4^2 \lambda _4+\lambda _{14} \left(\delta ^2+n_1^2\right)+\lambda _{24} \left(\delta
   ^2+n_2^2\right)+\lambda _{34} n_3^2\nonumber\\&&\qquad+\frac{\delta ^2 \left(n_1-n_2\right) \tilde{\lambda} _{24}}{n_1}\big]\,,\\
\tilde{\lambda} _{13}&=& -\frac{n_2 \tilde{\lambda} _{23}}{n_1} \,,  \\
\tilde{\lambda} _{14}&=& -\frac{n_2 \tilde{\lambda} _{24}}{n_1} \,,  \\ 
f&=& \frac{\delta}{\sqrt{2} k_0}\left[\left(n_2-n_1\right)
   \left(2 \lambda +\tilde{\lambda} _{12}\right)+\frac{n_2 \left(k_4^2 \tilde{\lambda} _{24}-n_3^2 \tilde{\lambda} _{23}\right)}{\delta ^2+n_1 n_2}\right]\,.\label{f}
\end{eqnarray}
Assuming that the bracketed factor in the rhs of Eq.(\ref{f}) is
nonvanishing, $\delta$ can be interpreted as an induced vev, whose
smallness is related to the scale of $\mathcal{L}$ symmetry breaking,
characterized by $f$, in the sense that in the limit $\delta\to 0$,
the symmetry of the potential is enhanced by $f \to 0$.

We conclude this section with an estimate of the scales involved and
the resulting light neutrino mass
\begin{equation}\label{numass2}
m_{light}\approx\frac{1}{\sqrt{2}}\frac{|y_1 \tilde{y}_2 \delta( n_1 -  n_2)|}{\sqrt{n_1^{2}y_1^{2}+n_2^{2}\tilde{y}_2^{2}}}.
\end{equation} 
Taking $f\sim\mathcal{O}(1)\,\mathrm{keV} $ ,
$k_0\sim k_4\sim\mathcal{O}(10^2) \,\mathrm{GeV} $,
$n_1\sim n_2\sim n_3\sim\mathcal{O}(1)\,\mathrm{TeV}$ and quartic
couplings of $\mathcal{O}(1)$ in Eq.(\ref{f}), the resulting scale
$\delta\sim\mathcal{O}(10)\,\mathrm{eV}$ can accommodate easily a
neutrino mass of $\mathcal{O}(10^{-1}) \,\mathrm{eV}$ without invoking
tiny Yukawa couplings or large values for the vacuum expectation
values $n_1\,,n_2$. Finally, note that the light neutrino mass in
Eq.(\ref{numass2}) is mostly insensitive to the values of $n_1$ and
$n_2$, and therefore new physics in this model can lie within reach of
the LHC experiments.

\section{Summary and discussion}
\label{sec:summary-discussion}

Here we have proposed a realistic \TrTrOne electroweak gauge model
with an enlarged Higgs sector.  The scheme leads to Dirac neutrinos,
and allows for the natural implementation of a type II seesaw
mechanism.
The model substantially improves the one previously considered in
Ref.~\cite{Valle:2016kyz}, in that here the small vacuum expectation
values associated to neutrino mass generation are decoupled from the
quark sector.
The charged lepton and quark masses are reproduced in a realistic way, avoiding the mixing between exotic and standard model quarks. The energy scales characterizing
neutrino mass generation and the new gauge interactions arising from
the enlarged \TrTrOne gauge symmetry could be accessible to the
current LHC experiments.

\section*{Acknowledgements}

This work is supported by the Spanish grants FPA2014-58183-P,
Multidark CSD2009-00064, SEV-2014-0398 (MINECO) and
PROMETEOII/2014/084 (Generalitat Valenciana). C.A.V-A. acknowledges
support from CONACyT (Mexico), grant 251357.

\bibliography{m331}

\end{document}